\documentclass[12pt,preprint]{aastex}

\slugcomment{Not to appear in Nonlearned J., 45.}

\shorttitle{Particle Acceleration in Turbulence}
\shortauthors{Matsukiyo \& Hada}

\begin{document}


\title{Relativistic particle acceleration in developing Alfv\'{e}n turbulence}


\author{S. Matsukiyo and T. Hada}
\affil{Department of Earth System Science and Technology, Kyushu University,\\
    6-1 Kasuga-Koen, Kasuga, 816-8580, Fukuoka, Japan}
\email{matsukiy@esst.kyushu-u.ac.jp}


\begin{abstract}
A new particle acceleration process in a developing Alfv\'{e}n turbulence 
in the course of successive parametric instabilities of a relativistic 
pair plasma is investigated by utilyzing one-dimensional electromagnetic 
full particle code. Coherent wave-particle interactions result in 
efficient particle acceleration leading to a power-law like energy 
distribution function. In the simulation high energy particles having 
large relativistic masses are preferentially accelerated as the 
turbulence spectrum evolves in time. 
Main acceleration mechanism is simultaneous relativistic resonance 
between a particle and two different waves. An analytical expression 
of maximum attainable energy in such wave-particle interactions is derived.
\end{abstract}





\section{Introduction}

Large amplitude Alfv\'{e}n waves are ubiquitous in space and astrophysical 
environments. Such waves are believed to play key roles in the so-called 
DSA (diffusive shock acceleration) process in which charged particles are 
diffusively accelerated in the course of multiple scattering through 
turbulent Alfv\'{e}n waves upstream and downstream of collisionless shocks. 
The DSA is widely accepted as one of the most efficient acceleration 
mechanisms of galactic cosmic rays of energy up to $\sim 10^{15.5} eV$ 
\citep{kry77,axf77,bel78,bla78,dru83,lag83,bla87,jon91,mal01,duf05}. 
In the process it is implicitly assumed that the Alfv\'{e}n turbulence 
is phase randomized and its spectrum is time stationary. On the 
other hand, turbulent Alfv\'{e}n waves commonly 
observed in the solar-terrestrial environments are often intermittent, 
and coherent MHD structures are frequently superposed. Namely, the 
phase random approximation cannot be assumed \citep{dew99,had03,nat06} 
and a spectrum of turbulence may evolve both in space and time \citep{bur05}. 
This is probably due to the fact that nonlinear wave-wave interactions 
tend to generate coherence among wave phases, and that spatial and temporal 
scales of relaxation processes in space plasmas are much larger than 
typical scales of our solar-terrestrial system. This may hold true 
with some high energy astrophysical environments. That is, spatial 
and temporal scales of a relaxation process are not negligibly small 
in comparison with scales of a whole acceleration site. Generally speaking, 
particle acceleration rate through coherent wave-particle interactions is much 
higher than that through incoherent, or fully turbulent, wave-particle 
interactions \citep{kur00,kur08}. Therefore, acceleration processes through coherent 
wave-particle interactions in a 'developing' turbulence, where a turbulence 
has not been fully developed, should be paid more 
serious attention. 

It is well-known that in various space plasma environments a nonequilibrium 
ion distribution function generates large amplitude Alfv\'{e}n waves 
through some instabilities and that those waves nonlinearly evolve to produce 
coherent magnetic wave forms. One of the common interpretations of generation 
mechanisms of such wave forms accompanying density fluctuations which are 
frequently observed in the solar wind \citep{spa97,spa88} is a 
parametric instability where nonlinear wave-wave 
interactions convert energy of a parent wave into several daughter waves 
with different frequencies and wavelengths \citep{gal63,sag69}. 
Although in the context of cosmic ray acceleration in turbulent media 
the above process was taken into account to estimate steady 
state distributions of wave intensities 
\citep{chi72,wen74,ski75a,ski75b,ski75c}, 
its developing processes and associated coherent wave-particle 
interactions have never been considered. Recent numerical studies 
revealed that successive parametric instabilities result in turbulent 
wave forms which have not been fully developed \citep{mat03,nar05,nar06}. 
However, only a few past studies 
paid much attention to particle acceleration processes in such a 
developing turbulence. Furthermore, 
only limited spatial and temporal evolutions of wave forms or spectra 
were discussed, since a computational resource was limited. 

In this paper long time evolution of parametric instabilities of a large 
amplitude Alfv\'{e}n wave in a rather large spatial domain is reproduced 
by utilizing one-dimensional relativistic full particle-in-cell (PIC) code. 
A plasma is assumed to be composed of electrons and positrons, since 
electron-positron pairs can be the dominant constituent of some high 
energy astrophysical plasmas like in the vicinity of a pulsar, 
active galactic nuclei (AGN), gamma ray burst (GRB), and so on. In the simulation 
we observe a particle acceleration process which is quite efficient 
and is due to interactions between coherent Alfv\'{e}n waves and 
relativistic particles. The process is quite different from some 
other coherent acceleration processes being discussed recently, in 
which high frequency electrostatic waves play essential roles, like electron 
surfing acceleration induced by a cross field Buneman instability 
\citep{sim00,mcc01,hos02,die04,die05,ama07} 
and wakefield acceleration \citep{taj79,kat83,lyu06,hos08}. 

Simulation settings and results are represented in section 2. 
The acceleration process is discussed in detail by using an analytical 
model and test particle simulation in section 3. Summary and discussions 
are given in section 4.

\section{1D PIC Simulation}

Long time evolution of parametric instabilities of a large amplitude 
monochromatic Alfv\'{e}n wave in a relativistic pair plasma is 
reproduced by performing a one-dimensional PIC simulation. A parent 
wave is given only at the beginning of the run as a monochromatic 
and right-handed circularly polarized Alfv\'{e}n wave with amplitude 
$B_p / B_0 = 1$ and wavenumber $k_0 c / \Omega_0 = 6.21$ (number of 
wave mode is 512) which gives a frequency $\omega_0 / \Omega_0 = 0.627$, 
where ${\bf B_0}$ is the ambient magnetic field which is along the 
$x-$axis, $c$ denotes speed of light, and $\Omega_0 = eB_0 / m_0 c$ 
is nonrelativistic gyrofrequency, respectively. Corresponding 
velocity perturbations of the pair plasma are given by the relativistic 
Walen relation \citep{had04}. Boundary conditions are periodic for 
particles and all field components. System size is $L = 518.1 c / \Omega_0$. 
Squared ratio of nonrelativistic gyro and plasma 
frequencies is $\Omega^2_0 / \omega^2_p = 0.1$, and normalized 
scalar temperature for both electrons and positrons is $T/m_0 c^2 = 
1.6 \times 10^{-3}$. 

\begin{figure}
\epsscale{1.0}
\plotone{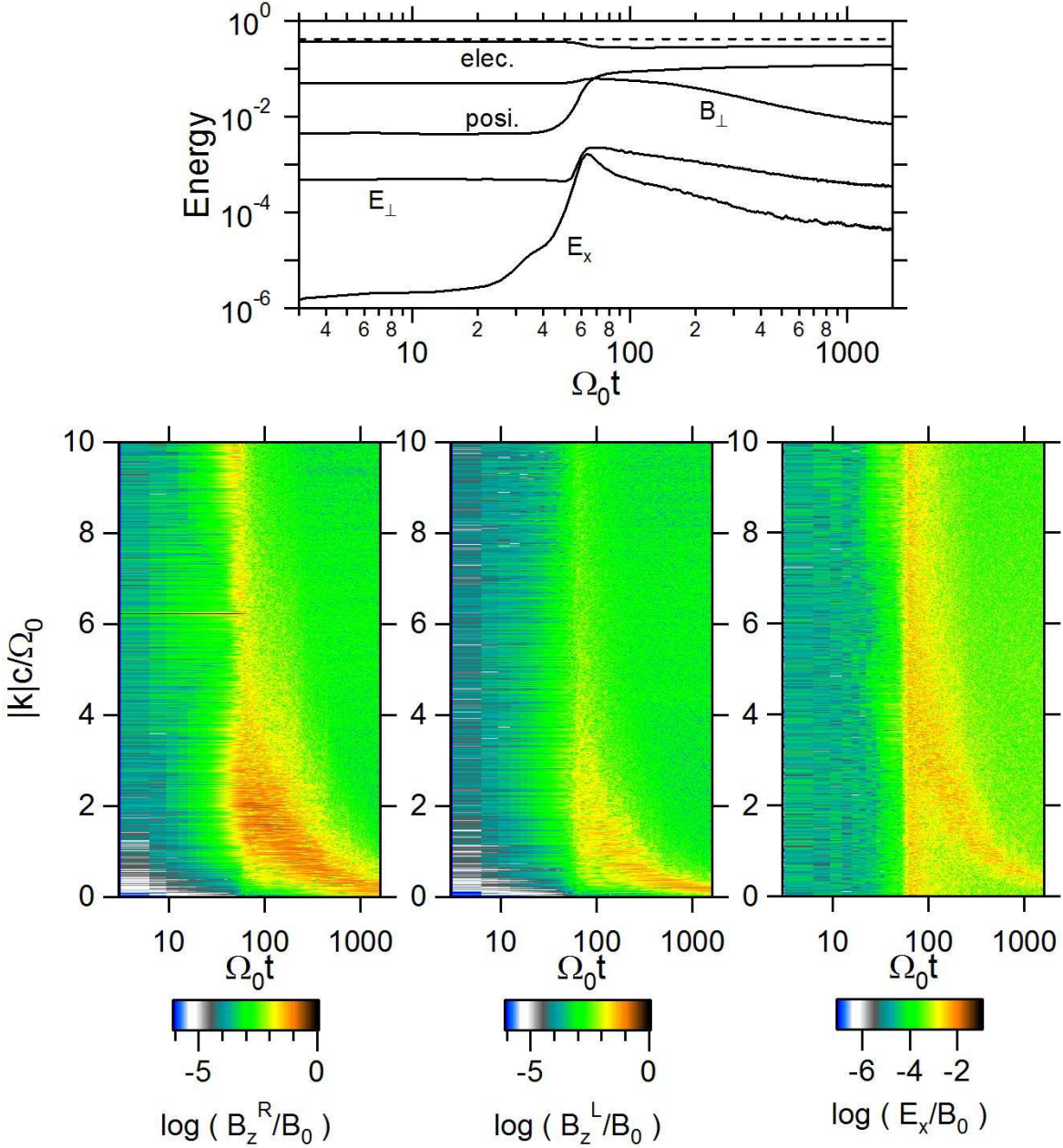}
\caption{Time evolution of wave power spectra. The right and left 
hand helicity modes of $B_z$ component are Fourier decomposed in left and 
middle panels, respectively. The right panel shows $E_x$ component. 
See the electronic edition of the Journal for a color version 
of this figure.\label{fig1}}
\end{figure}

Fig.\ref{fig1} shows time evolutions of energy densities (upper panel) and 
Fourier spectra of wave amplitudes for $B_z$ and $E_x$ field components (lower panels). 
In the upper panel the solid lines denote electron and positron kinetic 
energies (labeled by `elec.' and `posi.'), transverse magnetic field energy 
(`$B_{\perp}$'), and transverse and longitudinal electric field 
energies (`$E_{\perp}$' and `$E_x$'), respectively. The dashed 
line indicates the total energy, which is well conserved during the run. 
Although rather long system evolution (up to $\Omega_0 t = 1616$) is 
calculated, the system is still far from the so-called equilibrium state. 
In the lower panels wave amplitude spectra for 
positive and negative helicity (corresponding to positive and negative 
wavenumbers respectively) modes of the $B_z$ component are shown in the left ($B^R_z$) 
and middle ($B^L_z$) panels by using a technique of Fourier 
decomposition \citep{ter86}. Note that in the middle panel the actual 
sign of the wavenumber is negative. Since most of the daughter waves are 
right-hand polarized, the left (middle) panel shows wave intensity mainly of 
positively (negatively) propagating waves. At the beginning only the 
parent wave has significant intensity at $kc / \Omega_0 = 6.21$ in the 
left panel, although it is not particularly outstanding in the figure 
because of the narrow spectrum. 
In the early stage, $50 < \Omega_0 t < 100$, the parent wave energy is 
transfered to a variety of daughter waves through large number of 
channels of wave-wave interactions. 
In this stage two parametric instabilities are dominant. 
A modulational instability generates daughter magnetic fluctuations 
with the same signs of wavenumber as that of the parent wave. They are 
observed in the $B^R_z$ spectrum. Couplings among the daughter and 
the parent waves result in compressional electrostatic fluctuations 
with wavenumber smaller than that of the parent wave as seen in the 
right panel. In a decay instability, on the other hand, daughter 
magnetic fluctuations have negative wavenumbers which appear 
in the $B^L_Z$ spectrum, while daughter electrostatic 
fluctuations have wavenumbers larger than that of the parent wave in the 
right panel. Most of the parent wave energy is wasted in this stage. 
However, the intensities of the daughter waves are still large so that 
successive parametric instabilities occur thereafter. These successive 
processes are sustained mainly by decay instabilities which can be confirmed 
because peak wavenumbers of the $E_x$ spectrum is always larger than 
those of the $B^R_z$ and $B^L_z$ spectra. The previous simulation study 
by \citet{mat03} confirmed occurence of up to the second decay instability 
for the case of $B_p / B_0=0.1$.

Fig.\ref{fig2}a demonstrates electron energy distribution functions 
at various times. Fourier amplitude spectra of $B_z$ component at the 
corresponding times are plotted in Fig.\ref{fig2}b. 
Rapid heating occurs after the first instabilities developed ($\Omega_0 t = 95$: 
black dashed line). Up to this stage, most of the electrons show 
semi-stochastic motions in a noizy system with a primary wave. 
Here we refer the primary wave and the noize to the parent wave and 
superposition of the daughter waves, respectively. This results in rapid scattering 
in pitch angle as shown in Fig.\ref{fig2}c, where an average pitch angle 
of electrons with respect to ${\bf B_0}$ and 
its standard deviation are plotted as a function of time. 
The standard deviation just after its rapid growth at $\Omega_0 t \sim 70$ 
is $\sim 0.5$, which roughly coincides with an analytical estimate of the 
maximum pitch angle width of an electron resonating with the monochromatic 
parent wave (see Appendix). 
After the parent wave disappears, most of the electrons are detrapped by 
the parent wave and start to wander in the phase space. 
As time passes, a high energy tail appears in the electron distribution 
function (Fig.\ref{fig2}a), which gradually approaches a power-law 
spectrum with index $\alpha \sim -2.5$ 
($\Omega_0 t = 316$: gray solid line, $1616$: black solid line). 
At the same time the wave amplitude spectrum develops the power-law type 
spectrum also, with the index $\alpha \sim -2.0$, while the wave spectral 
peak shifts toward lower wavenumbers.

A behaviour of the most efficiently accelerated electron is shown in Fig.\ref{fig3}. 
Fig.\ref{fig3}a indicates energy time history of the electron. Spatial trajectory 
of the electron during $0 < \Omega_0 t < 600$ ($800 < \Omega_0 t < 1400$) 
is plotted in Fig.\ref{fig3}c(d). A background color scale denotes strength of 
magnetic fluctuations $B_{\perp} \equiv \sqrt{B^2_y + B^2_z}$. 
Horizontal arrows in Fig.\ref{fig3}c(d) and Fig.\ref{fig3}a correspond to each other 
denoting time intervals where strong accelerations occur. Fig.\ref{fig3}e(f) 
shows a snapshot of the wave profile at the time denoted as the dashed line 
in Fig.\ref{fig3}c(d). The red, blue, 
and black lines represent $B_y$, $B_z$, and an envelope, respectively. 
The dotted line indicates the position of the electron. It is seen that 
strong acceleration occurs when the electron is trapped in a trough 
of the magnetic envelope. Such sharp envelope troughs are temporarily observed 
in various regions of the system throughout the run. Hence, the electron 
experiences similar acceleration processes several times. In this example 
one may recognize four such acceleration periods. Only the second 
and the third periods are marked (the first (fourth) one is 
$160 < \Omega_0 t < 190$ ($\Omega_0 t > 1450$)). 
The acceleration occurs always perpendicular to ${\bf B_0}$ as seen 
in Fig.\ref{fig3}b which shows a trajectory of the electron in the 
$u_{\perp}-u_{\parallel}$ space, where $u_{\perp}$ and $u_{\parallel}$ 
denote the four-velocities perpendicular and parallel to ${\bf B_0}$. 
The red markers 
indicate positions of the electron in the phase space at the same time 
as Fig.\ref{fig3}e (lower marker) and Fig.\ref{fig3}f (upper marker). 
We checked a hundred most efficiently accelerated electrons' trajectories 
and confirmed that all of them show essentially the same features as mentioned 
above, i.e., trapping within the envelope troughs and successive perpendicular 
acceleration. In the next section the acceleration process is modeled and analyzed 
in detail.

\section{Acceleration of High Energy Electrons}

\subsection{Coherent waves observed in the simulation: modeling}

In the PIC simulation shown above sharp magnetic envelope troughs 
are locally formed throughout the period of strong electron accelerations. 
In a successive decay process a number of daughter Alfv\'{e}n waves 
propagating both parallel and antiparallel to ${\bf B_0}$ are excited. 
Depending on their phases, amplitudes of some wave modes are sometimes 
locally canceled out each other, or they are simply less dominant in 
amplitude. Then, there appear some regions where two oppositely 
propagating waves dominate. 
Here, the envelope structures observed in the PIC simulation are modeled 
by a superposition of such two oppositely propagating waves as follows. 
\begin{equation}
\label{Bw}
  \left(
  \begin{array}{c}
    B^y_w \\
    B^z_w \\
  \end{array}
  \right)
  = B_w \cos kx 
  \left(
  \begin{array}{c}
    {\cos \omega t}\\
    {\sin \omega t}\\
  \end{array}
  \right) \hspace{7mm}
\end{equation}
\begin{equation}
\label{Ew}
  \left(
  \begin{array}{c}
    E^y_w \\
    E^z_w \\
  \end{array}
  \right)
  = -{\omega \over kc} B_w \sin kx 
  \left(
  \begin{array}{c}
    {\cos \omega t}\\
    {\sin \omega t}\\
  \end{array}
  \right)
\end{equation}
Eq.(\ref{Bw}) is equivallent to 
$$
  {\bf B_w} = {B_w \over 2} 
  \left(
  \begin{array}{c}
    {\cos(kx - \omega t)}\\
   -{\sin(kx - \omega t)}\\
  \end{array}
  \right)
  + {B_w \over 2}
  \left(
  \begin{array}{c}
    {\cos(-kx - \omega t)}\\
   -{\sin(-kx - \omega t)}\\
  \end{array}
  \right).
$$
Hereafter, we assume $\omega$ and $k$ are both positive without losing 
generality. Typical waveforms at two different time domains are shown in 
Fig.\ref{fig4}a and b. The black solid and dashed lines denote $y$ 
and $z$ components of rotating carrier waves, and the gray lines 
represent envelopes which is independent of time. In the neighborhood 
of a trough, it is confirmed that 
eqs.(\ref{Bw}) and (\ref{Ew}) give a reasonable model of a waveform 
observed in the PIC simulation shown in Fig.\ref{fig4}c. In the following 
motion of a test particle in this system is analyzed.

\subsection{Motion of a particle in given EM fields}

In this section we consider motion of an electron in the electromagnetic 
waves given by eqs.(\ref{Bw}) and (\ref{Ew}). The equation of motion is 
\begin{equation}
\label{eqom}
  {d{\bf u} \over dt} = -{e \over m_0 c} \left( {\bf E} + {{\bf u} \over 
  \gamma} \times {\bf B} \right),
\end{equation}
\begin{equation}
\label{dxdt}
  {dx \over dt} = {u_x \over \gamma} c,
\end{equation}
where ${\bf u} \equiv \gamma {\bf v} / c$, $\gamma = \sqrt{1+u^2}$, 
$e$ and $m_0$ indicate charge and rest mass of the electron, 
${\bf B} = B_0{\bf x} + {\bf B_w}$, and ${\bf E} = {\bf E_w}$, respectively. 
When we write ${\bf u} = (u_{\parallel}(t), u_{\perp}(t) \cos\phi(t), 
u_{\perp}(t) \sin\phi(t))$ and introduce normalized variables as 
$\xi=x \Omega_0 / c$, $\tau=\Omega_0 t$, $\kappa = kc / \Omega_0$, 
$\nu = \omega / \Omega_0$, $b_w = B_w / B_0$, and $v_{ph} = \nu / \kappa$, 
eqs.(\ref{eqom}) and (\ref{dxdt}) are written 
as follows.
\begin{equation}
\label{dux}
  \dot{u}_{\parallel} = b_w {u_{\perp} \over \gamma} 
  \cos \kappa \xi \sin\psi
\end{equation}
\begin{equation}
\label{duperp}
  \dot{u}_{\perp} = b_w \left( v_{ph} \sin \kappa \xi \cos\psi 
  - {u_{\parallel} \over\ \gamma} \cos \kappa \xi \sin\psi \right),
\end{equation}
\begin{equation}
\label{dpsi}
  \dot{\psi} = -{b_w \over u_{\perp}} \left( v_{ph} \sin \kappa \xi 
  \sin\psi + {u_{\parallel} \over\ \gamma} \cos \kappa \xi \cos\psi \right)
  - \left( \nu - {1 \over \gamma} \right).
\end{equation}
\begin{equation}
\label{dxi}
  \dot \xi = {u_{\parallel} \over \gamma}.
\end{equation}
Here, $\psi = \phi - \nu \tau$, and the dot denotes time derivative 
in terms of $\tau$, respectively. Note that a variation of normalized 
particle energy, or the Lorentz factor, is given as 
\begin{equation}
\label{dgam}
  \dot{\gamma} = b_w v_{ph} 
  {u_{\perp} \over \gamma} \sin \kappa \xi \cos\psi.
\end{equation}
In the following behaviours of the electron is discussed by using 
eqs.(\ref{dux})-(\ref{dgam}).

\subsection{Fixed point analysis}

The above set of equations clearly has several fixed points as listed in 
Table \ref{tbl1}. Here, $u_0$ denotes a value of $u_{\perp}$ at a corresponding fixed 
point and $\gamma_0 = \sqrt{1+u^2_0}$. Apparently, the fixed points I - III 
(IV - VI) correspond to troughs (crests) of the magnetic envelope. As far as the 
acceleration is concerned, it is easily inferred that the fixed points IV - VI 
are not important since the electric field strength is very weak around 
there. Actually, efficient acceleration observed in the PIC simulation 
always occurs around the troughs of the magnetic envelope. Therefore, 
only the fixed points I - III are focused here. Stability of these fixed 
points is discussed in the following.

\subsubsection{Stability of the fixed points I - III}

Expanding eqs.(\ref{dux})-(\ref{dgam}) around the fixed point I and 
retaining only the first order terms, we obtain 
\begin{equation}
\label{delua}
  \delta \dot{u}_\parallel = -b_w {u_0 \over \gamma_0} \kappa \delta \xi,
\end{equation}
\begin{equation}
\label{delue}
  \delta \dot{u}_\perp = -b_w v_{ph} \delta \psi,
\end{equation}
\begin{equation}
\label{delpsi}
  \delta \dot{\psi} = {b_w \over u^2_0} v_{ph} \delta u_{\perp} - 
  {1 \over \gamma^2_0} \delta \gamma,
\end{equation}
\begin{equation}
\label{delx}
  \delta \dot{\xi} = {\delta u_{\parallel} \over \gamma_0}
\end{equation}
\begin{equation}
\label{delgam}
  \delta \dot{\gamma} = -b_w {u_0 \over \gamma_0} v_{ph} 
  \delta \psi = {u_0 \over \gamma_0} \delta \dot{u}_{\perp},
\end{equation}
where $\delta$ denotes small first order quantities.

From eqs.(\ref{delua}) and (\ref{delx}), we have 
\begin{equation}
\label{ddua}
  \delta \ddot{u}_{\parallel} = -b_w {u_0 \over \gamma^2_0} \kappa \delta 
  u_{\parallel}.
\end{equation}
The above expression represents a trapping oscillation with 
trapping frequency $\omega_{trap} = (b_w u_0 \kappa / \gamma^2_0)^{1/2}$, 
which is rewritten with the original parameters as 
$\omega_{trap} = (B_w kc u_0 / B_0 \Omega_0 \gamma^2_0)^{1/2}$. 

From eqs.(\ref{delue}), (\ref{delpsi}), and (\ref{delgam}), on the other hand, 
we obtain 
\begin{equation}
\label{ddue}
  \delta \ddot{u}_{\perp} = {b_w u_0 v_{ph} \over \gamma^3_0} 
  \left( 1 - b_w v_{ph} {\gamma^3_0 \over u^3_0} \right) \delta u_{\perp}.
\end{equation}
Here, the first term in the parenthesis (or the second term in the right 
hand side of 
eq.(\ref{delpsi})) arises from the variation of the Lorentz factor, i.e., 
due to the relativistic effect. In the non-relativistic limit, therefore, 
eq.(\ref{ddue}) represents a trapping motion in the perpendicular momentum 
space with frequency $\omega_{trap} = b_w v_{ph} / u_0 = (B_w/B_0) 
(\omega / kc) (1 / u_0)$. 
However, in the relativistic case with $b_w v_{ph} \gamma^3_0 / u^3_0 < 1$, 
$\delta u_{\perp}$ (and $\delta \psi$) 
diverges in time. When $u_0$ is small enough so that $\gamma_0 \approx 1$, 
the above inequality is hardly satisfied. In such a case, the fixed point 
is stable. But if $u_0$ becomes large and the inequality is satisfied, 
such an electron gains transverse energy while it keeps being trapped in the 
$x-$direction. We consider this solution later more in detail.

For the fixed point II, equations corresponding 
to eqs.(\ref{ddua}) and (\ref{ddue}) are obtained by formally changing 
the sign of $b_w$. Therefore, the system is unstable for parallel 
fluctuations while it is stable for perpendicular fluctuations. 
This fixed point is actually conjugate to the relativistic fixed point 
discussed above. Another fixed point (III), conjugate to the nonrelativistic 
one, which should be a saddle point in $u_{\perp}-\psi$ phase space, 
appears at $(u_{\perp}, \psi) = (0, n \pi)$ as singular points of $d u_{\perp} / 
d \psi$, where $n = 0, \pm1, \pm2, \cdots$.

\subsubsection{Features in $u_{\perp}-\psi$ phase space}

In order to study perpendicular dynamics let us first 
consider a reduced system in which parallel quantities are 
fixed at the fixed point, i.e., $u_{\parallel} = 0$ and $\kappa \xi = \pi/2$. 
Then we only have to consider 
the following two equations.
\begin{equation}
\label{reddu}
  \dot p = 2 \sqrt{p} b_w v_{ph} \cos \psi
\end{equation}
\begin{equation}
\label{redpsi}
  \dot \psi = -{b_w \over \sqrt{p}} v_{ph} \sin \psi - 
  \left( \nu - {1 \over \sqrt{1+p}} \right)
\end{equation}
Here, $p=u^2_{\perp}$ has been introduced. 
This system has a Hamiltonian defined as
\begin{equation}
\label{hamilton}
  H(p, \psi) = 2 \sqrt{p} b_w v_{ph} \sin \psi + 
  \nu p - 2 \sqrt{1+p},
\end{equation}
where $\dot p = \partial H / \partial \psi$ and $\dot \psi = - 
\partial H / \partial p$ are satisfied. Contours of the Hamiltonian 
for $b_w = 1.0$, $v_{ph} = 0.17$, and $\nu = 0.11$ are represented as 
gray lines of the upper (linear scale) and lower panels (logarithmic scale) 
in Fig.\ref{fig5}a. The above parameters 
are chosen so that the waves interacting with the accelerated electron in 
the second acceleration stage observed in the PIC simulation (Fig.\ref{fig3}e) is 
appropriately 
reproduced. It is easily confirmed that a center at $\psi = \pi/2$, 
which is clearly recognized in the lower panel, corresponds to the 
nonrelativistic stable fixed point of the fixed point I and another center at 
$\psi = 3\pi/2$, which can be seen both in the upper and lower panels, 
to the relativistic fixed point II discussed above. 

The nonrelativistic center should satisfy the following condition from 
Table \ref{tbl1}. 
\begin{equation}
\label{const}
  -{b_w v_{ph} \over u_0} = \nu - {1 \over \gamma_0}
\end{equation}
Although $\nu-1/\gamma_0=0$ in the small amplitude limit, it is never satisfied 
in the nonrelativistic case since $\nu < 1$. Therefore, 
the nonrelativistic center appears only when wave amplitude becomes 
finite. The closed trajectories around this center essentially 
coincides with the 
nonresonant trapping discussed by \citet{kur05}. 

At the relativistic center, the resonance condition $\nu-1/\gamma_0=0$ 
should be satisfied in the 
ultrarelativistic limit, since $b_w v_{ph} / u_0$ is negligible. In such a
 case relativistic decrease of the gyro 
frequency allows another resonance with low frequency waves. This 
resonance can also be present in small amplitude limit. Relativistic 
linear resonance conditions between an electron and two oppositely 
propagating waves ($\gamma \nu \mp \kappa u_{\parallel} - 1 = 0$) 
are shown in Fig.\ref{fig5}c. 
The two resonance lines intersect at $u_{\parallel}=0$ where 
$\nu - 1/\gamma = 0$ and $\gamma = \sqrt{1 
+ u^2_{\perp}}$ are satisfied, while there never appears such an 
intersection in nonrelativistic limit. This indicates that an electron 
can resonate simultaneously with two waves in the relativistic case. 
By using eq.(\ref{hamilton}), the maximum width of the separatrix in 
$u_{\perp}$ is estimated as 
\begin{equation}
\label{width}
  \Delta u_{\perp} = 4 \sqrt{{b_w \over \kappa \nu}}.
\end{equation}
Furthermore, the maximum $u_{\perp}$ on the separatrix is given by
\begin{equation}
\label{maxu}
  u_{\perp, max} = \left( \sqrt{{b_w \over \kappa}} + 
  \sqrt{{1 \over \nu}} \right)^2 
  = \left( \sqrt{{B_w \over B_0}{\Omega_0 \over kc}} + 
  \sqrt{\Omega_0 \over \omega} \right)^2.
\end{equation}
For instance, $u_{\perp, max} \approx 18.4$ for the parameters used in Fig.\ref{fig5}.
This gives a good agreement with the electron energy achieved in the 
end of the second acceleration stage observed in the PIC simulation. 

So far we have only looked at a subset of the system in which 
$\delta u_{\parallel} = 0$. If we allow $\delta u_{\parallel}$ to slightly 
deviated from zero, it should satisfy 
\begin{equation}
\label{dua}
 \delta \ddot{u}_{\parallel} = -b_w {u_0 \over \gamma^2_0} 
 \kappa \sin \psi \delta u_{\parallel}
\end{equation}
around the fixed point in the original system eqs.(\ref{dux})-(\ref{dxi}). 
Hence, the system is stable for $0 < \psi < \pi$ and is unstable for 
$\pi < \psi < 2\pi$ in terms of parallel fluctuations. 
The black solid line in Fig.\ref{fig5}a and \ref{fig5}b shows a numerical solution of 
eqs.(\ref{dux})-(\ref{dxi}). A trajectory of the electron initially 
positioned near the separatrix at $\psi=0$ (indecated by a small arrow 
in Fig.\ref{fig5}a) is represented. The electron 
moves alomost along the separatrix when it is in $0 < \psi (< \pi)$. 
But when the electron enters in $\pi < \psi < 2\pi$, its trajectory 
starts deviating from the separatrix. And finally it is clearly detrapped when it 
approaches $\psi = 3 \pi / 2$. The trajectory in 
the momentum space shown in Fig. \ref{fig5}b is very similar to what is 
observed in the PIC simulation which is again plotted as a gray line.

\subsection{Statistics of high energy particles}

Let us briefly discuss the power-law like distribution of electorns in the current 
acceleration process. It is shown in Fig.\ref{fig2}a that a high energy tail with 
power-law index $\sim -2.5$ evolves in the late stage of the simulation. 
This power-law nature may be related with power-law spectrum of 
wave amplitude of magnetic fluctuations. In the stage of successive decay instabilities 
($\Omega_0 t > 100$) a spectral cascade of the magnetic fluctuations 
occurs with keeping a power-law index of the wave amplitudes almost constant with 
$\sim -2.0$ (Fig.\ref{fig2}b). The enhancement of the wave amplitude at higher 
wavenumber regime, $kc/\Omega_0 > 3$, is the remnant of initial parent 
wave and side band waves generated by the modulational instabilities 
in the early stage of the run. In order to confirm correlations between 
the wave and electron energy spectra, the following test particle simulation 
with periodic boundary conditions is performed. Waves are given by 
superposition of a number of right hand 
polarized Alfv\'{e}n waves with power-law spectrum shown in Fig.\ref{fig6} 
(power-law index $\alpha$ is an external parameter), which models the lower wavenumber 
part of Fig.\ref{fig2}b. Both positive and negative wavenumber modes are 
evenly distributed. The wave form is given by 
\begin{equation}
\label{modelb}
  \left(
  \begin{array}{c}
    B_y \\
    B_z \\
  \end{array}
  \right)
  =\sum^3_{kc/\Omega_0=-3} b_w B_0 \left| {k \over K_0} \right|^{\alpha} 
  \left(
  \begin{array}{c}
    \sin(k(x-v_{ph}t)+\Phi_k) \\
    \cos(k(x-v_{ph}t)+\Phi_k) \\
  \end{array}
  \right),
\end{equation}
where $v_{ph}/c = 0.17$ and $\{ \Phi_k \}$ are initial wave phases which are randomly 
distributed at $t=0$, and corresponding transverse electric fields are 
given by ${\bf k} \times {\bf E} = (|k| v_{ph} / c) {\bf B}$. The power-law 
index $\alpha$ is fixed to 1.0 for $k_{min} \le k \le K_0$, 
where $k_{min} \equiv 2 \pi / L$ and $K_0$ is the coherence wavenumber. 
The system size $L = 80 \pi c/\Omega_0$ is common for all the following runs, 
and $b_w = 0.75$ and $K_0 c / \Omega_0 = 0.1$ if not specified.

In such a system where a number of waves propagate in two opposite directions, 
particles are expected to be stochastically accelerated via second order 
Fermi like process. First, this is confirmed by solving nonrelativistic 
equations of motion of $10^5$ electrons. An initial distribution function 
is given as a spatially homogeneous gyrotropic ring distribution with 
$v_{\perp}/v_{ph} = 0.1$. Fig.\ref{fig7} shows distributions in 
(a) $v_{\perp} - v_{\parallel}$ and (b) $X-W_{kin}$ phase spaces, and 
(c) energy distribution functions at two different times, 
$\Omega_0 t = 1000$ and $3000$, respectively, where 
${\bf v}=(v_{\parallel}, v_{\perp})$ denotes particle velocity and 
$W_{kin} = v^2/2 v^2_{ph}$ is normalized kinetic energy. 
The solid gray lines in Fig.\ref{fig7}b 
denote envelope profiles of given magnetic fluctuations. The 
particles are spatially bunched around the troughs of the 
magnetic envelopes because of mirror effect. In the 
velocity space the particles rapidly pitch angle diffuse 
(Fig.\ref{fig7}a) while the averaged particle 
energy slowly increases in time (Fig.\ref{fig7}c). 
These are the features of a stochastic or a second order Fermi acceleration. 
In comparison, drastic changes occur if relativistic effects 
are taken into account. The simulation results with the same 
initial conditions as in the nonrelativistic case are plotted in 
Fig.\ref{fig8} with the same format as Fig.\ref{fig7}. Here, the normalized kinetic 
energy is defined as $W_{kin} = (\gamma - 1) c^2 /v^2_{ph}$. 
Compared with the nonlerativistic 
case, the maximum particle energies at each corresponding time are extremely 
higher, and they are at the same order as that obtained from 
eq.(\ref{maxu}) with $k=K_0$ and $b_w=0.75$ (the black solid and black dashed 
lines in Fig.\ref{fig8}c). 
Spatial bunching of particles similar to the nonrelativistic case 
is found in Fig.\ref{fig8}b, while a fraction of the bunched particles are 
accelerated to extremely high energies. Such high energy particles have rather large 
pitch angles, denoting perpendicular acceleration (Fig.\ref{fig8}a). 
Interestingly, a power-law like energy distribution appears only 
at $\Omega_0 t = 1000$ in Fig.\ref{fig8}c, although the corresponding 
energy range in the distribution function is small (the black solid line). 
The power-law index does not change 
even in the case with a different initial ring velocity, 
$v_{\perp} / v_{ph} = 5.1$ (not shown). However, the spectrum becomes softer 
when $\alpha=-3.0$ is assumed (the gray dashed line in Fig.\ref{fig8}c). 
When the coherence wavenumber is large as $K_0 c / \Omega_0=0.4$, 
the bulk electrons are accelerated and the energy distribution is 
no longer power-law at $\Omega_0 t = 1000$ (the gray dotted line in Fig.\ref{fig8}c). 
It is noted from eqs.(\ref{width}) and (\ref{maxu}) that the minimum 
$u_{\perp}$ of the separatrix of the relativistic resonance 
of the dominant wave mode at $k=K_0$ decreases with increasing $K_0$. 
Hence, most of electrons which initially distribute below the separatrix 
can enter inside the separatrix through stochastic motions at rather 
early stage, and they can be perpendicularly accelerated within a short 
time in the similar way seen in the previous section (cf. Fig.\ref{fig5}). 
The peak energy is roughly 
consistent with the $u_{\perp, max} (\sim 940)$ estimated by using $b_w=0.75$ 
and $\kappa=0.4$ from eq.(\ref{maxu}). 
In contrast, for small wave amplitude, $b_w=0.1$, the maximum energy at 
$\Omega_0 t = 1000$ is much smaller than the value obtained from 
eq.(\ref{maxu}) (not shown). The reason may be that in this run the 
minimum $u_{\perp}$ of the separatrix is rather high because of its 
narrow width so that particles which initially distribute far below the 
separatrix have not entered in it until this time.

In Fig.\ref{fig8}c the power-law like nature appears as a transient state in 
the system at $\Omega_0 t = 1000$ for $\alpha = -2.0$. In this run, at 
later time ($\Omega_0 t = 3000$), high energy end of the
particle distribution is so enhanced that the energy distribution does 
not fit the power-law spectrum(black dashed line). It is confirmed that the 
hump of the high energy part grows at least till $\Omega_0 t = 6000$. 
After sufficiently long time, it probably results in bulk acceleration 
as seen in the case of small $K_0$. In the PIC 
simulation the wave spectrum 
is not time stationary but cascading through the successive decay 
instabilities as already mentioned. In other words, a wave with 
a certain wavenumber has finite life time. This may be why the high 
energy tail evolves without extra accumulation at high energy end 
of the distribution function in the late stage, $\Omega_0 t > 316$, 
of the PIC simulation (Fig.\ref{fig2}a). To confirm this, an additional 
test particle simulation is performed by assuming that wavenumber 
of the magnetic field with maximum intensity varies in time as 
$K_0 c / \Omega_0 \sim (\Omega_0 t)^{-0.9}$ which 
gives a reasonable fit in the late stage of the PIC simulation. 
The particle distribution and waves at $\Omega_0 t = 1000$ in the 
run corresponding to the black solid line in Fig.\ref{fig8}c are 
chosen as initial conditions. Then, in the later time $\Omega_0 t = 3000$ 
high energy tail extends roughly obeying the power-law and some low 
energy particles remain unaccelerated, as shown in Fig.\ref{fig8}c as the 
gray solid line.

\section{Summary and Discussions}

In the present paper an efficient particle acceleration process in the 
course of successive parametric instabilities of large amplitude Alfv\'{e}n 
waves was investigated. The acceleration takes place as a result of 
interactions between coherent waves in the developing Alfv\'{e}n 
turbulence and relativistic particles. An important point is that 
relativistic wave-particle interactions allow simultaneous resonance 
between a particle and two different waves. The maximum attainable energy 
through this acceleration process was analytically estimated. 

In this acceleration process a high energy particle 
is preferentially accelerated. During the successive decay instabilities, 
a peak of the wave Fourier spectrum shifts in time toward a lower frequency 
(longer wavelength) regime. Because of relativistic decrease of particle's 
gyro frequency, low frequency waves preferentially resonate with and 
accelerate particles with large energy. Therefore, if once a particle is 
accelerated and 
its effective mass is increased, in later time such a particle easilt 
satisfies the resonance condition with lower frequency waves generated by 
successive decay processes. Fig.\ref{fig3}a shows an example of such an electron's 
energy time history in which one can recognize four acceleration phases 
around $\Omega_0 t \sim 170, 430, 1200$, and $1550$.

By utilizing test particle simulations in section 4, it is shown that relativistic 
effetcs are essential in producing nonthermal particles. Note that the system 
discussed in section 3 is exactly the same as that discussed by 
\citet{wyk01a,wyk01b}, although they mainly focused on stochastic 
off-resonance diffusion in the phase space with an application to planetary 
magnetospheres in mind. Actually, the regular trajectories in Fig.3(d) in 
\citet{wyk01a} correspond to the relativistic resonance focused in 
this paper. Enhancement of Lyapunov exponent at $\omega \sim \Omega_0 / n \gamma$, 
where $n$ is an integer, shown in their Fig.8 might be related with 
the relativistic resonance. In our test particle simulation 
the power-law like energy distribution function similar to what was 
observed in the PIC simulation is reproduced by assuming the time 
evolving Fourier spectrum of Alfv\'{e}n turbulence. Finite life time 
of a wave mode in developing turbulence may contribute 
to creation of such a distribution function. 
However, details of the acceleration process including analytical estimate 
of the power-law index are still unclear and will be investigated 
in the near future.

The acceleration process discussed in this paper is essentially different 
from some of the recently studied coherent acceleration processes, i.e., electron 
surfing acceleration and wakefield acceleration. In these processes 
electrostatic field plays essential roles. Furthermore, the processes mainly 
affect electrons since generated electrostatic waves have rather 
high frequencies, while ion acceleration may also occur in a very late 
stage of nonlinear evolution of a system \citep{hos08}. 
In the acceleration process discussed here roles of electrostatic fields 
are subdominant, although they are necessary for the decay instabilities. 
It should be noted further that the process may be able to act on ions too 
when a left hand polarized wave is introduced as an initial parent wave. 
The Alfv\'{e}n waves are essentially incompressible so that they 
may survive for rather long period compared with electrostatic Langmuir 
or ion acoustic waves. Hence, the process may last for long time, and 
may also follow the above mentioned electrostatic 
acceleration processes in some occasions.

In the present study all the multidimensional effects have been excluded. 
In higher dimensional cases daughter waves propagating in oblique to the 
ambient magnetic field can also have finite growth rates \citep{vin91}. 
It is confirmed by MHD simulation that these waves destroy the planar 
structure assured in a one dimensioal simulation \citep{gho93,gho94,del01}. 
However, in low frequency regime the decay instability of parallel 
progagation is dominant \citep{vin91}. Therefore, at least in such a regime 
the acceleration process observed here are expected to work, while 
acceleration rate may decrease to some degree because of appearance 
of nonplanar structures. This is similar to what is discussed for 
electron surfing acceleration in which the acceleration takes place 
even in two dimensional cases despite decrease of acceleration rate 
\citep{ama08}. At all events, further investigations are necessary for estimate 
efficiency of this acceleration process in a more realistic situation.

We finally give a comment on applications of this acceleration process. Since 
the situation simulated in section 2 is rather general, there may  be several 
fields of application like a pulsar wind nebula, a pulsar magnetosphere, an 
outflow of GRB, and an AGN jet, and so on. Vicinity of a relativistic shock 
is probably one of the candidates. Similary to the earth's foreshock, 
large amplitude Alfv\'{e}n waves may be generated by beam-plasma interactions 
between backstreaming ions and an upstream plasma, and the waves nonlineary 
evolve via parametric instabilities. Since the acceleration is efficient and 
locally takes place, it may contribute to the so-called injection process 
into the DSA. As another possible case, cosmic ray-plasma interactions 
upstream of a supernova remnant shock are now extensively studied after 
the pioneering works by \citet{luc00}, \citet{bel01}, and \citet{bel04,bel05}. 
Most of such studies pay attention to amplification 
of upstream magnetic fluctuations that can be scatterers of cosmic rays. 
And that is expected to result in increase of maximum 
attainable energy in the DSA process. On the other hand, amplified magnetic 
fluctuations may lead to the local and the coherent acceleration 
process discussed here. In other words, some particles may be accelerated 
through this process without crossing a shock. In this sense the process is 
similar to the so-called second order Fermi acceleration, although an 
efficiency of the acceleration process discussed here is much higher than that of 
the second order Fermi process as shown by the test particle simulation.

\acknowledgments

The authors thank Victor Mu\~{n}oz for useful discussions. The PIC simulation 
was performed by the super computer in ISAS/JAXA Sagamihara. This work was 
supported in part by Incentive aid to the prominent research, 
Interdisciplinary graduate school of engineering sciences, Kyushu University 2007.




\appendix

\section{Motion of a relativistic particle in a monochromatic wave}

\begin{figure}
\epsscale{1.0}
\plotone{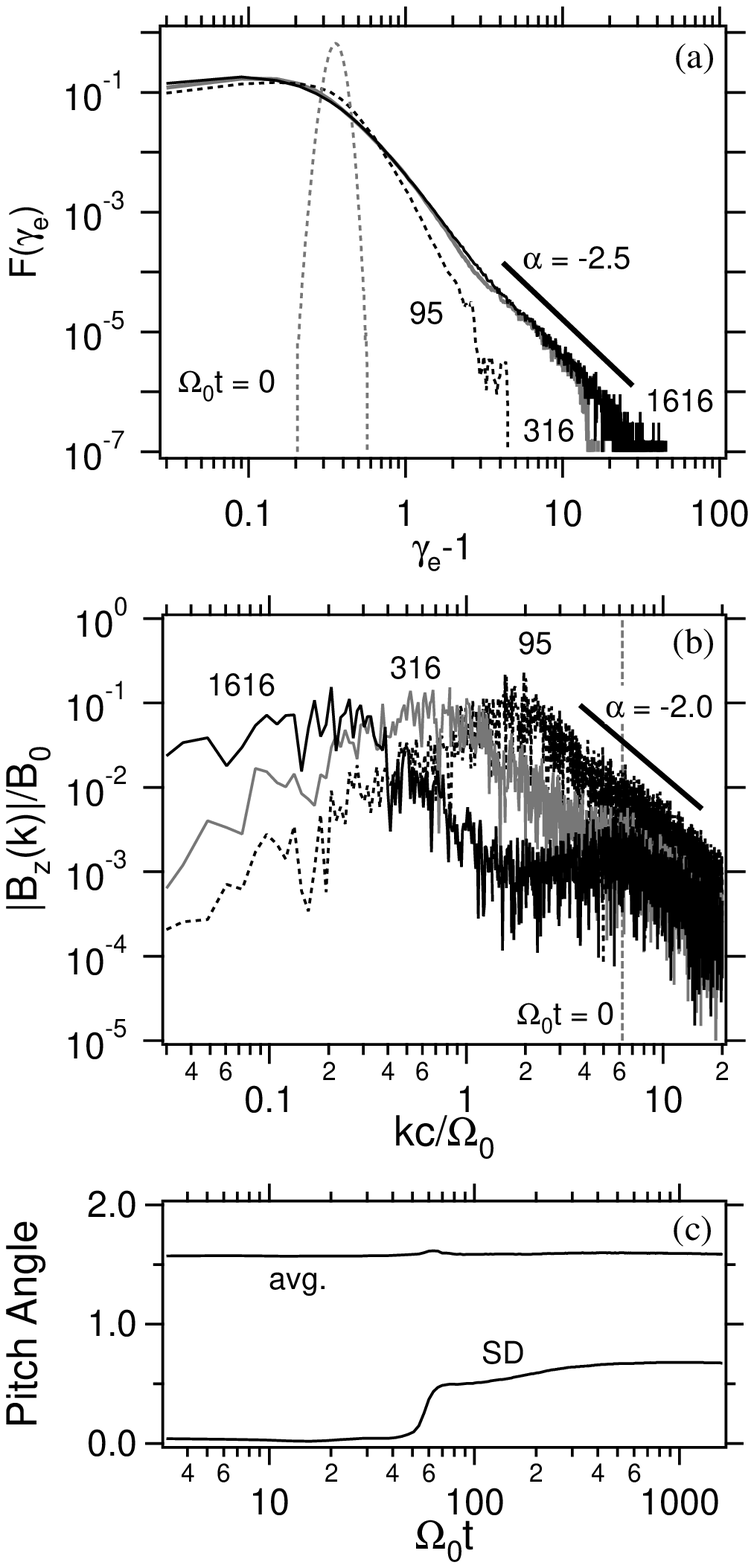}
\caption{Electron energy distribution functions at various times.\label{fig2}}
\end{figure}

\begin{figure}
\epsscale{1.0}
\plotone{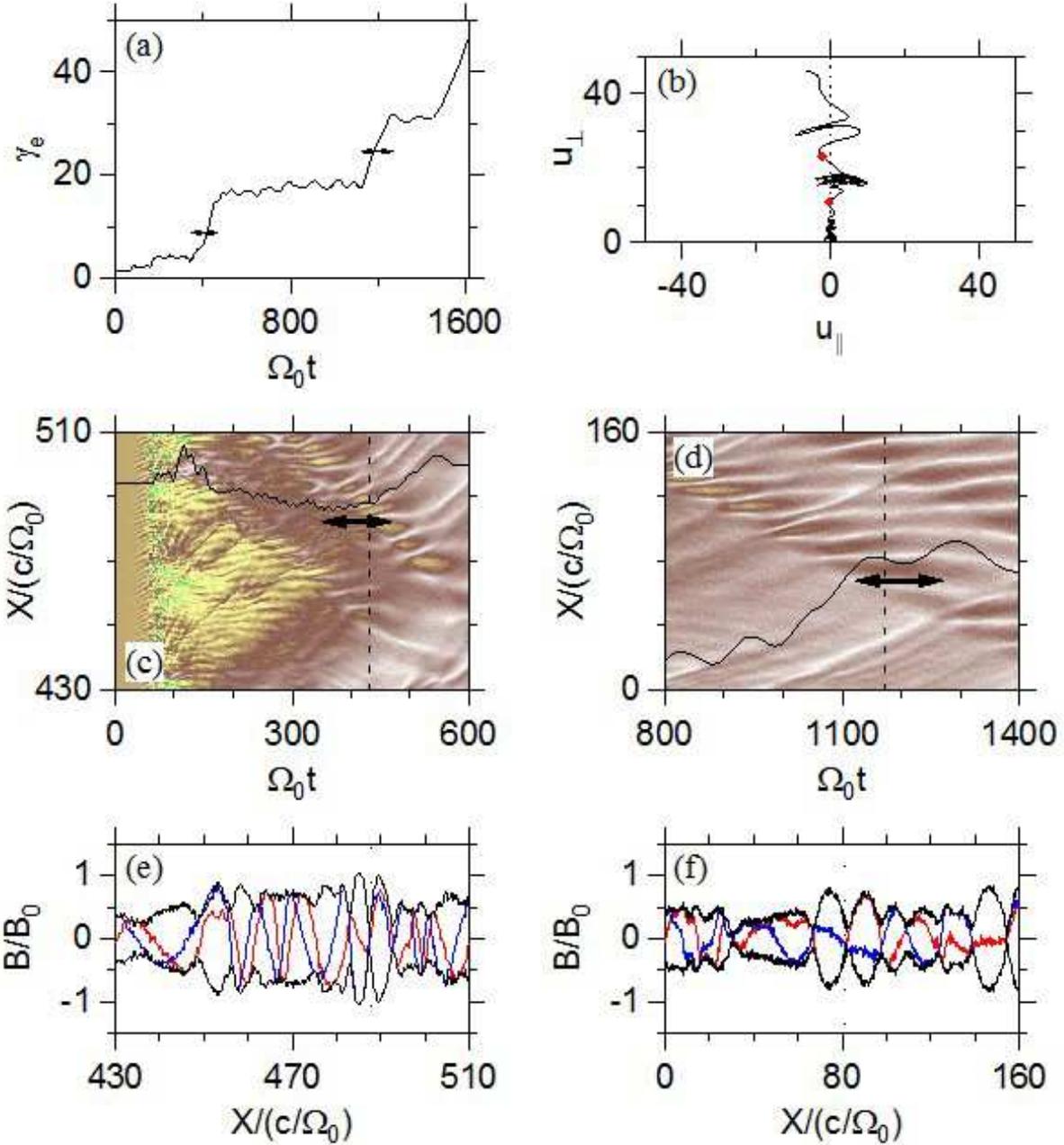}
\caption{Behaviours of an accelerated electron. (a) Time history 
of the electron energy. (b) Trajectories of the electron in $u_{\perp}-
u_{\parallel}$ space and (c)(d) in $X-t$ space. The arrows in (c) and 
(d) indicate time domains where strong acceleration occur as shown by 
the arrows in (a). Background color scale in (c) and (d) denotes 
amplitude of magnetic fluctuations. (e)(f) Spatial profiles of magnetic 
fluctuations in two acceleration phases represented by dashed lines in 
(c) and (d). The red, blue, and black solid lines show $B_y$, $B_z$ 
components and envelope, respectively. The dotted lines show positions 
of the particle at correspponding times. 
See the electronic edition of the Journal for a color version 
of this figure.\label{fig3}}
\end{figure}

\begin{figure}
\epsscale{1.0}
\plotone{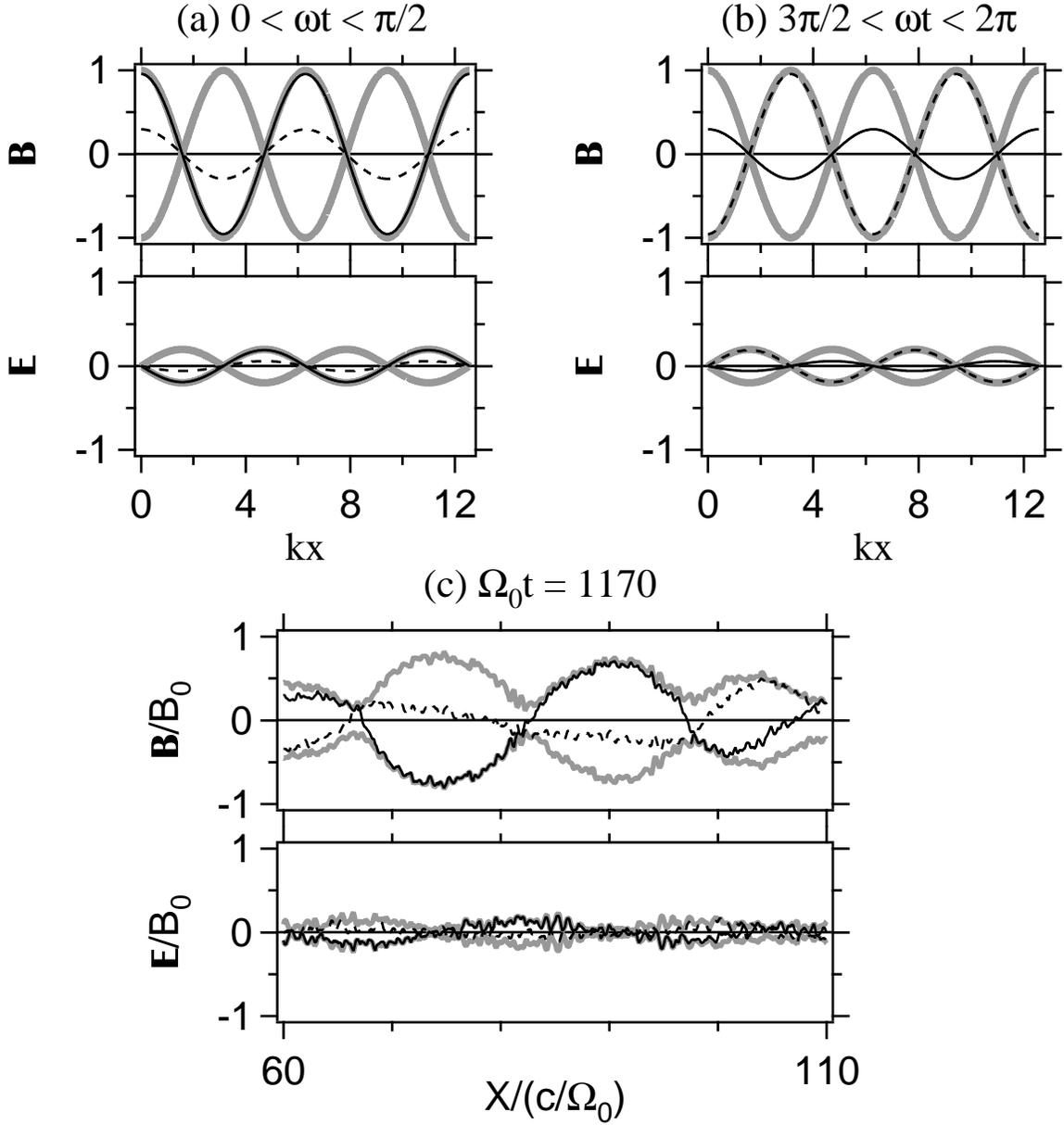}
\caption{(a), (b) Modeled wave forms at different time domains ($B_w = 1, 
\omega / kc = 0.2$), and (c) an example of a waveform seen in the PIC 
simulation at $\Omega_0 t = 1170$.\label{fig4}}
\end{figure}

\begin{figure}
\epsscale{1.0}
\plotone{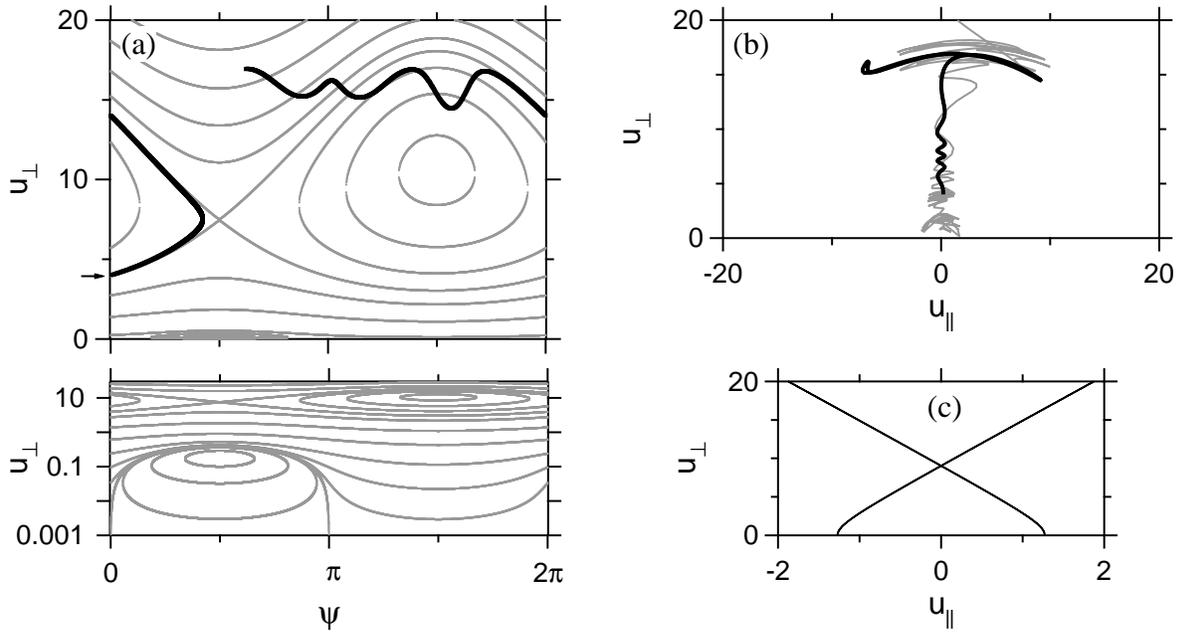}
\caption{Electron trajectories in (a) $u_{\perp}-\psi$, 
(b) $u_{\perp}-u_{\parallel}$ phase space, and (c) linear resonance 
conditions of wave-particle interactions in relativistic (solid line) 
and nonrelativistic (dashed line) plamsas. 
Solutions of eq.(\ref{hamilton}) are plotted as gray lines in (a). 
In (b) a gray line denotes a trajectory of the electron shown in 
Fig.\ref{fig3}.  A numerical solution of eqs.(\ref{dux}) - (\ref{dxi}) for 
$\Omega_w / \Omega_0 = 1, kc/|\Omega_0| = 0.65$, and 
$\omega / |\Omega_0| = 0.11$ is indicated as black lines 
in (a) and (b).\label{fig5}}
\end{figure}

\begin{figure}
\epsscale{1.0}
\plotone{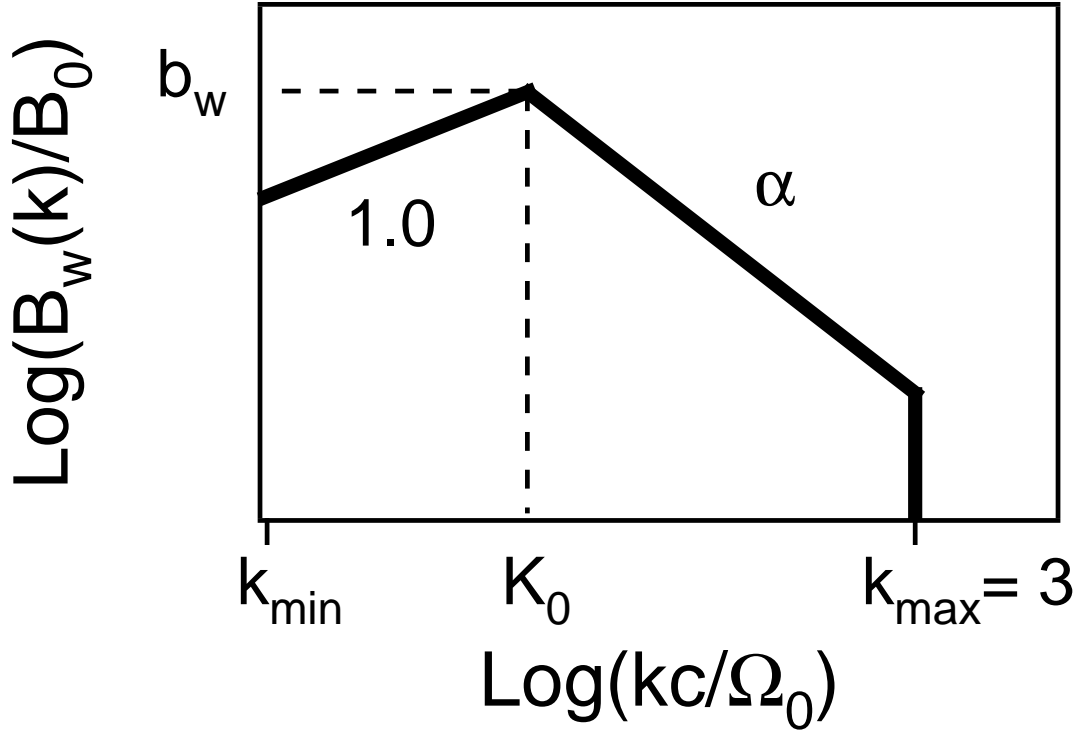}
\caption{Wave power spectrum used in 
the test particle simulation.\label{fig6}}
\end{figure}

\begin{figure}
\epsscale{1.0}
\plotone{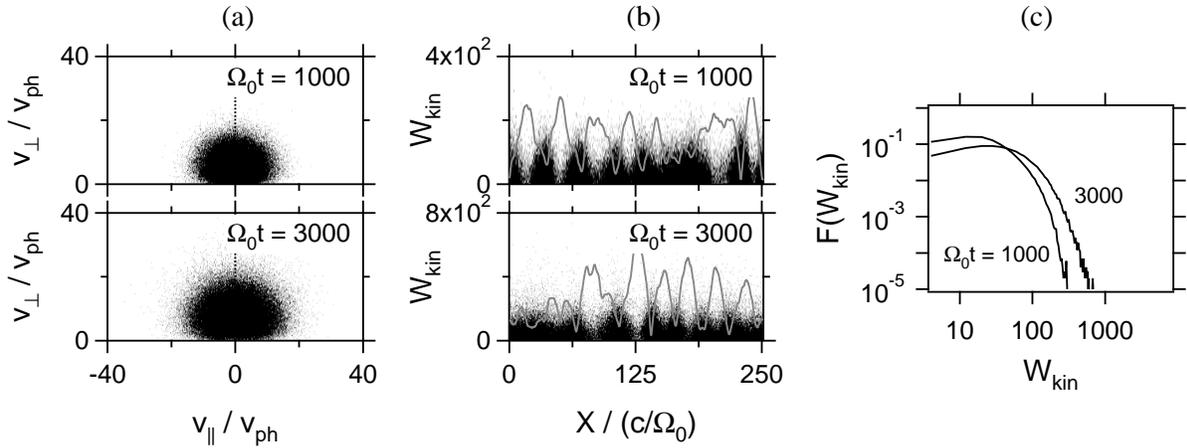}
\caption{Results of nonrelativistic test particle simulation. 
Electron distributions in (a) $v_{\perp}-v_{\parallel}$ and 
(b) $X-W_{kin}$ phase spaces, and (c) energy distribution functions at 
$\Omega_0 t = 1000$ and $3000$, respectively. The solid gray lines in 
(b) denote profiles of magnetic field envelopes at the corresponding times.
\label{fig7}}
\end{figure}

\begin{figure}
\epsscale{1.0}
\plotone{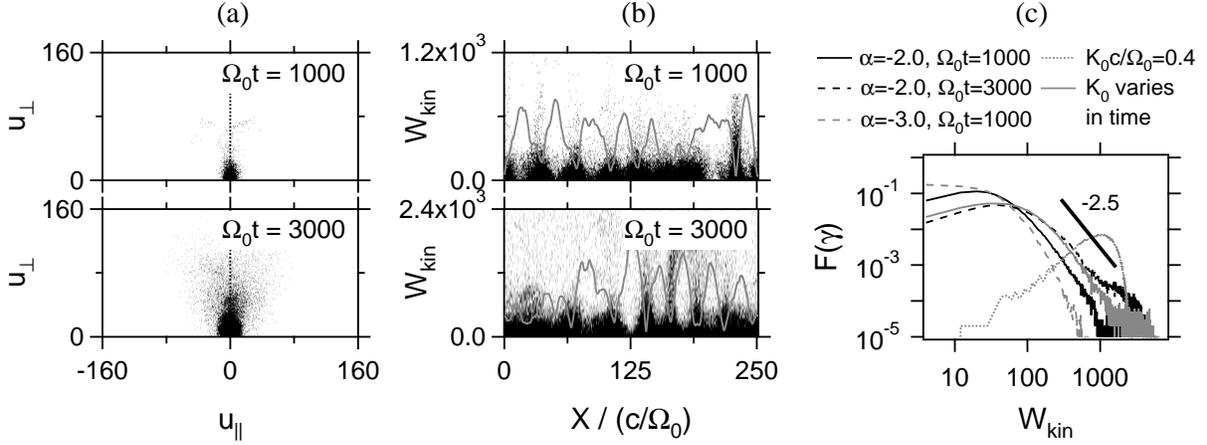}
\caption{Results of relativistic test particle simulation in the same 
format as Fig.\ref{fig7}. A number of energy distribution functions 
are plotted in (c). See datails in the text.
\label{fig8}}
\end{figure}

\begin{figure}
\epsscale{1.0}
\plotone{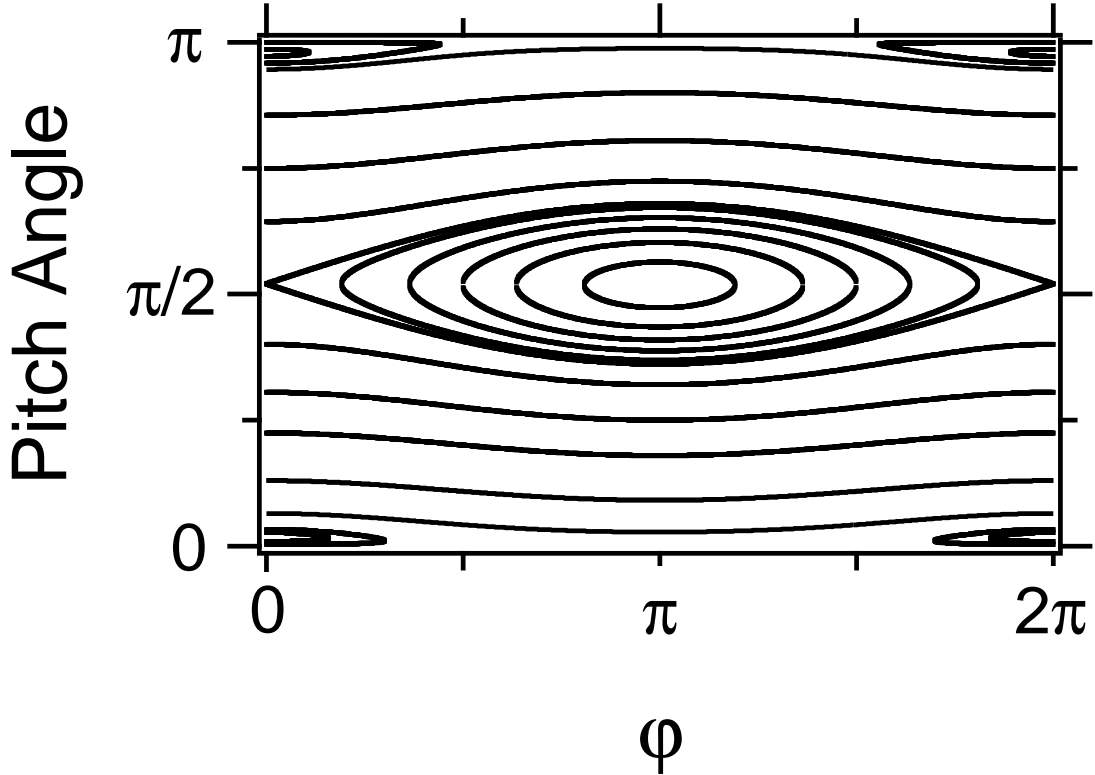}
\caption{Trajectories of relativistic electrons interacting with the monochromatic 
circularly polarized parent wave. $B_w / B_0 = 1$, $kc / \Omega_0 = 6.21$, 
and $u = 0.64$ are assumed.
\label{figa}}
\end{figure}

According to \citet{kur05}, an equation of motion of an electron in a 
monochromatic circularly polarized wave in a wave frame is reduced as 
\begin{equation}
  \dot \mu = - {\partial H \over \partial \psi},\hspace{2mm} \dot \psi = 
{\partial H \over \partial \mu},
\end{equation}
where $\mu$ is the pitch angle cosine, $\psi$ the particle gyrophase with 
respect to the wave phase, and
\begin{equation}
  H = {K \over 2} \left( \mu + {1 \over K} \right)^2 + b_w 
  \sqrt{1 - \mu^2} \cos \psi
\end{equation}
is Hamiltonian. The above expressions are held even when relativistic effects 
are taken into account by putting $K = ukc / \Omega_0$ and $b_w = (B_w / B_0) 
\sqrt{1 - v^2_{ph} / c^2}$. Electron trajectries interacting with the 
parent wave ($B_w / B_0 = 1$ and $kc/\Omega_0 = 6.21$) for different values of $H$ 
are plotted in Fig.\ref{figa}. Here, $u=0.64$ is assumed since the value is 
derived as average one at $\Omega_0 t = 70$ in the PIC simulation. The factor 
of $v^2_{ph} / c^2$ is neglected because of its smallness. Fig.\ref{figa} 
is essentially the same as Fig.5 in \citet{kur05}. The maximum value of half 
the width of the separatrix is $\sim 0.5$ (at $\psi=\pi$), which roughly 
coincides with the standard deviation of pitch angle observed in the PIC 
simulation at $\Omega_0 t = 70$ (Fig.\ref{fig2}c).

%
%

\clearpage

\begin{table}
\begin{center}
\caption{Fixed points\label{tbl1}}
\begin{tabular}{c|ccccc}
\tableline\tableline
     & $\kappa \xi$ & $\psi$ & $u_{\parallel}$ & $u_{\perp}$ & constraint\\
\tableline
  I & $\pm {\pi \over 2}$ & $\pm {\pi \over 2}$ & 0 & 
  $-{b_w v_{ph} \over u_0} = (\nu - {1 \over \gamma_0})$ &\\
  II & $\pm {\pi \over 2}$ & $\mp {\pi \over 2}$ & 0 & \hspace{3mm}
   ${b_w v_{ph} \over u_0} = (\nu - {1 \over \gamma_0})$ &\\
  III & $\pm {\pi \over 2}$ & $0(\pi)$ & 0 & 0 & ${\sin \psi \over u_0} 
   \rightarrow \mp (\nu - {1 \over \gamma_0})/b_w v_{ph}$ \\
  IV & $0(\pi)$ & $0(\pi)$ & 0 & $\nu - {1 \over \gamma_0} = 0$ &\\
  V & $0(\pi)$ & $\pi(0)$ & 0 & $\nu - {1 \over \gamma_0} = 0$ &\\
  VI & $0(\pi)$ & $\pm {\pi \over 2}$ & 0 & 0 & ${\sin \kappa \xi \over u_0} 
   \rightarrow \mp (\nu - {1 \over \gamma_0})/b_w v_{ph}$ \\
\tableline
\end{tabular}
\end{center}
\end{table}

\clearpage

%

%

%

%

%

%

%

%

\end{document}